# Exoplanet Analog Observations of Earth from Galileo Disk-integrated Photometry

Ryder H. Strauss[1], Tyler D. Robinson[1,2,3,4], David E. Trilling[1], Ryan Cummings[1], and Christopher J. Smith[1]
[1] Department of Astronomy and Planetary Science, Northern Arizona University, Box 6010, Flagstaff, AZ 86011, USA; rhs72@nau.edu
[2] Lunar and Planetary Laboratory, Department of Planetary Sciences, University of Arizona, Tucson, AZ 85721, USA
[3] Habitability, Atmospheres, and Biosignatures Laboratory, University of Arizona, Tucson, AZ 85721, USA
[4] NASA Nexus for Exoplanet System Science Virtual Planetary Laboratory, University of Washington, Seattle, WA 98195, USA

## Abstract

The Galileo spacecraft had distant encounters with Earth in 1990 and 1992. Limited Solid State Imager (SSI) data acquired during these encounters has been previously presented, but the majority of the data from these Earth flybys have not been presented in the literature. Observations of Earth taken from afar are both rare and directly relevant to the development of any future exo-Earth direct imaging mission. Here we present a pipeline that vets, calibrates, and measures the disk-integrated brightness of the Earth, in multiple filters, from the complete SSI data sets from both the 1990 and 1992 Galileo flybys. The result is over 1500 usable photometric measurements for Earth as an analog for an exoplanet. The 1990 data set includes full rotational lightcurves in six bandpasses spanning the optical range. The 1992 data set is more limited, with lightcurves only spanning 14 hr. Time-averaged photometry for both encounters is presented while variability and color are discussed relative to findings from NASA's EPOXI mission (which also provided photometric lighturves for Earth). The new Galileo/SSI data are used to further validate the Virtual Planetary Laboratory 3D spectral Earth model, which often serves as a stand-in for true disk-integrated observations of our planet. The revived Galileo/SSI data for Earth is a testament to the ability of NASA's Planetary Data System to maintain data over decades-long timescales. The disk-integrated products derived from these data add to a very short list of calibrated and published whole-disk observations of the Pale Blue Dot.

*Unified Astronomy Thesaurus concepts:* Photometry (1234); Exoplanet detection methods (489); Direct imaging (387)

## 1. Introduction

Whole-disk observations of solar system worlds provide a unique bridge between solar system planetary science and exoplanetary science (Kane et al. 2021). Not only do these data transform well-proven instrumentation for solar system remote sensing into new vehicles for exoplanet exploration, but the resulting data sets can prove crucial in validating approaches to exoplanet characterization (Marley et al. 2014; Lupu et al. 2016; Mayorga et al. 2016; Heng & Li 2021; Tribbett et al. 2021; Robinson & Salvador 2022). Foremost among solar system targets for exoplanet analog observations is Earth—the only known inhabited world and the archetypal habitable planet that would be characterized by under-development space telescope missions (Gaudi et al. 2018; Quanz et al. 2018; Roberge & Moustakas 2018), including the recently dubbed "Habitable Worlds Observatory" recommended by the Decadal Survey on Astronomy and Astrophysics for the 2020s (National Academies of Science, Engineering, and Medicine 2023).

Unfortunately, whole-disk observations of the distant Earth that are most suitable as exoplanet analog data are quite rare (Robinson & Reinhard 2020). Such observations should be acquired at a sufficient range that the local horizon reaches the planetary limb, which presents challenges to the use of data from low-Earth or geosynchronous orbit (Hearty et al. 2009). Earthshine observations—which record Earth light reflected from the portion of the Lunar disk unilluminated by the Sun (Galilei 1632; Danjon 1928; Dubois 1947)—have proven useful (Goode et al. 2001; Woolf et al. 2002; Pallé et al. 2003; Qiu et al. 2003; Seager et al. 2005; Montañés-Rodríguez et al. 2006; Turnbull et al. 2006), but can be challenging to calibrate and are limited to spectral ranges where Earth's atmosphere is sufficiently transparent. Data from interplanetary spacecraft are the best analog for exoplanet observations, but reduced and published results remain sparse (Christensen & Pearl 1997; Fujii et al. 2011; Livengood et al. 2011; Robinson et al. 2014; Yang et al. 2018).

Perhaps the most famous study of the distant Earth from mission data comes from Sagan et al. (1993), who used spatially resolved flyby observations from the Galileo spacecraft to explore pathways for life detection. Crucial data were provided by the Solid State Imager (SSI; Belton et al. 1992) and the Near-Infrared Mapping Spectrometer (NIMS; Carlson et al. 1992). Earth observations from NIMS are also explored in (Drossart et al. 1993), but suffer from widespread saturation issues. The original work from Sagan et al. (1993) does not report the total volume of SSI data for Earth, and the presentation of only a few images might lead a reader to believe that the data sets from the 1990 and 1992 flybys are sparse. In contrast, a query of NASA's Planetary Data System for SSI observations of Earth returns thousands of image files that could be processed into a more comprehensive data set for the Pale Blue Dot.

The work presented below details the creation of a pipeline to convert raw Galileo/SSI images of Earth into calibrated data products, including disk-integrated photometry. As directly imaged exoplanets are viewed as point sources, disk-integrated views of the distant Earth are the most relevant to observational studies of Earth in the context of exoplanets. The spatially integrated observations are then used to create lightcurves and

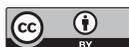

**Table 1**
Table of Key Information for 1990 and 1992 Galileo Earth Data Sets

| Epoch | Observation Start (UT) | Observation End (UT) | Midtime Phase Angle (deg) | Range (km) | Total # | Usable # |
|---|---|---|---|---|---|---|
| 1990 | 1990-12-11 14:06:29.722 | 1990-12-12 15:18:06.394 | 35.7 | 2,500,000 | 1820 | 1406 |
| 1992 | 1992-12-16 15:10:00.420 | 1992-12-17 05:05:11.063 | 88.3 | 6,300,000 | 3894 | 166 |

**Note.** Phase angle is computed from the mean value for the usable images in each data set. Observation start and end times denote the UT time of the first and last usable exposure.

broadband spectra. These processed products are compared to observations from NASA's EPOXI mission (Livengood et al. 2011), which also acquired time-resolved, disk-integrated photometry. Finally, the new Galileo data set is discussed within the context of existing observations of the Pale Blue Dot and ongoing mission development efforts.

## 2. Data Reduction

The juNASA PDS Imaging Node maintains a Galileo/SSI data set for Earth that consists of 5714 unique file pairs of raw data (stored as binary `IMG` files) and associated labels (stored as ASCII `LBL` files). Details on the two data sets available from the two Earth encounters are provided in Table 1. The USGS Integrated Software for Imagers and Spectrometers (`ISIS`[5]) software package—which contains legacy calibration information for SSI data for each Earth flyby—was used to reduce the SSI data images. The `glssi2isis` ISIS routine was used to convert the `IMG` files into `ISIS` cube files, which would be compatible with additional ISIS functions. The radiometric calibration was performed using `gllssical`, one such function which calibrates for dark current in the CCD and applies the wavelength-dependent conversions from pixel counts to spectral radiance. Here, each individual pixel has a spectral radiance value with units of $nW\,cm^{-2}\,nm^{-1}\,sr^{-1}$. Finally, the `isis2fits` routine was used to produce calibrated, spatially resolved `FITS` files. These files could then be used within standard Python-based image analysis software packages.

Most of our 5714 SSI images were of limited utility for exoplanet analog study due to either saturation or resolution/pointing issues. The calibrated `FITS` files use a standard value for saturated pixels, and roughly 1000 images were rejected from the 1990 and 1992 data sets for saturating over more than 10% of the total image. Spacecraft attitude information stored in the metadata for the images enabled the identification of cases where either: (1) Earth's disk was larger than the instrument field of view; or (2) the SSI instrument was not appropriately pointed at Earth. The distance between the spacecraft and the Earth at the time of observation was used to determine whether the entire disk was visible; if the Earth's projected diameter was larger than the SSI field of view in a given image, that image would be culled. Other attitude parameters in the metadata were used to determine if Earth was well centered in the image. The `SUB_SPACECRAFT_LINE` and `SUB_SPACECRAFT_LINE_SAMPLE` fields within the `FITS` headers indicate the approximate pixel position of the center of the target, and if either of these parameters were less than the previously computed radius away from the image edge, the images were culled. This removed most images in which the Earth was spilling off one side of the image. The subspacecraft pixel estimate was used to check for appropriate pointing. These quality checks removed a further 3548 images from the combined data sets.

Some additional processing of the images was necessary before photometry could be performed. A five-pixel-wide mask was applied to all four edges of each image to account for artifacts along the chip boundaries. Additionally, for those images which contained saturated pixels but did not exceed the threshold noted above, those saturated pixel values were replaced with an average value calculated by the nearest nonsaturated pixel in each direction.

An automated photometry pipeline was developed and included the placement of an aperture over Earth's disk in each image, thereby minimizing background ("sky") noise contributions. A perfect circle is a good approximation of the shape of the disk as the ellipticity of Earth is less than the pixel precision of the SSI images, so circular apertures were used for all of the photometry. The first step of this pipeline involved an automated edge detection using the Canny edge detection technique (Canny 1986). This approach produces an image array with nonzero values only along the edges of structure within the image. Edge detection could be impacted by cloud and surface features within Earth's disk, so aggressive thresholding was used to bring the true edge of the disk into relief. Figure 1 shows an example result when applying Canny edge detection to a SSI Earth image.

At this point, the images were ready for disk detection. As seen in Figure 1, edge detection provides a moderate-quality estimate but lacks accuracy. The disk edges are not smooth, and some structure from the interior survives the aggressive thresholding. Additionally, because the nightside terminator is not a sharp line, it is position in the edged image is highly dependent on the choice of threshold. Instead, this step of the pipeline made use of a circle Hough transform, which is a computer vision image processing technique used to algorithmically detect circles within an image (Xie & Ji 2002). The transform works by building a three-dimensional parameter space spanned by the three dimensions describing a circle: the position of the circle's center in the horizontal dimension ($a$), the vertical dimension ($b$), and the circle's radius ($r$). The image pixels are then randomly sampled and for every nonzero pixel, a surface is drawn in the parameter space with every {$a$, $b$, $r$} value that satisfies the circle equation. The highest density of nonzero points within the three-dimensional parameter space can be treated as the best-fit values for the circle in the data. Figure 2 demonstrates a two-dimensional slice in the Hough circle parameter space described here.

The known separation between Galileo and Earth, as well as Earth's known radius, was used to reduce the dimensionality of the Hough transform, thereby minimizing computational demand. Given the known angular size of Earth in each image, the only remaining free parameters are the horizontal and vertical positions of the disk center. This reduced the parameter space to two-dimensions, which improves both runtime and the

---
[5] https://isis.astrogeology.usgs.gov/7.0.0/index.html

**Table 2**
Time-averaged Photometry of Earth from the 1990 and 1992 Galileo Encounters

| Epoch | VIOLET | GREEN | RED | IR-7270 | IR-7560 | IR-8890 | IR-9680 |
|---|---|---|---|---|---|---|---|
| 1990 | 80.00265 ± 0.00004 | 79.5406 ± 0.00006 | 63.2551 ± 0.00003 | 50.8580 ± 0.00003 | 43.6600 ± 0.00003 | 37.0359 ± 0.00002 | 23.6362 ± 0.00001 |
| 1992 | 65.70 ± 1.20 | 43.50 ± 0.54 | 38.40 ± 0.54 | 35.70 ± 1.35 | 23.40 ± 0.45 | 54.00 ± 1.86 | 85.50 ± 1.14 |

**Note.** The 1990 photometry is reported as the rotation-averaged specific intensity in each available filter. The 1992 photometry is reported as the specific intensity in the best individual images from each available filter.

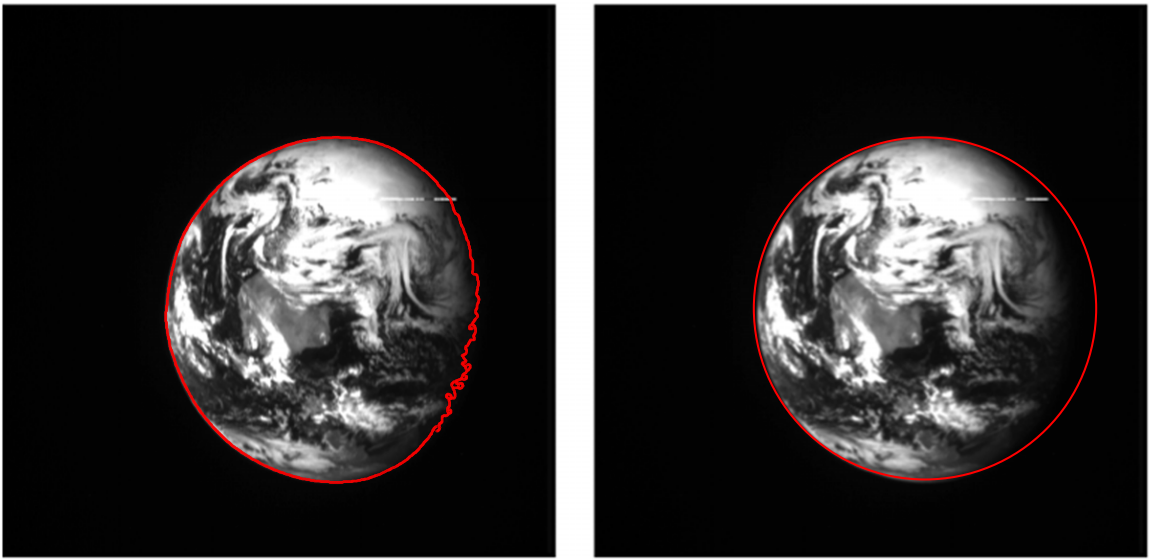

**Figure 1.** Left: result of thresholded Canny edge detection (red) on a resolved SSI image of Earth. Note the irregularity along the nightside terminator in the edge detection. Crucially, the location of the edge detected here is not absolute. Instead, its position is dependent on the threshold applied. A higher threshold would push this edge closer to the center of the disk, and a lower threshold would do the opposite. Thus, a more robust disk detection is necessary. Right: image of Earth with a correct Hough transform identification of disk edge shown in red.

reliability of the disk detection. Disk detections were verified both through spot checking individual images as well as inspecting animations of the 1990 and 1992 encounters (see Figure 1).

Standard aperture photometry techniques (Mighell 1999) were applied to the images following successful disk detection. Because the data were already calibrated to radiometric units on a per-pixel level, the pixel counts within the aperture in each image were averaged to produce distance-independent specific intensity values. A "sky" background subtraction is also applied, in which the median "sky" pixel value is subtracted from each pixel in the image. Finally, photon counting errors and the root-mean-square scatter in background pixel counts were combined to yield the overall photometric errors on the disk-integrated brightness. The resulting computed uncertainties, when also scaled by the distance to Earth, match well to the perceived scatter in the lightcurves.

### 3. Results

Table 2 summarizes the time-averaged photometry from the Galileo data. Figures 3 and 4 show the calibrated lightcurves (in intensity units) for the 1990 and 1992 epochs, respectively. The 1990 data provides very dense coverage of a full 24 hr rotation in six (of eight) different SSI filters. (Data in 1990 were not acquired in the SSI IR-8890 and white-light/clear filters.) The 1992 data is much more limited, with lightcurve data in only three filters, a lower overall data quality (due in part to a larger separation between the spacecraft and Earth), and coverage spanning only 14 hr. While both sets of lightcurves do indicate significant color-dependent variability, some analyses below focus only on the significantly more complete 1990 data set.

Figure 5 shows time-averaged spectrophotometry for the 1990 and 1992 Earth encounters. The longer duration of the 1990 data set enables the full rotational averaging of the photometric data. For the 1992 data set, though, averaging is only over the spanned 14 hr of useful data. Noisy, single-image photometry from the 1992 encounter is provided for the GREEN, RED, IR-7560, and IR-8890 filters, which were not shown in the earlier lightcurve results owing to the extremely sparse nature of data collection in these filters for this epoch. A reflectivity measure was created by dividing-out the incident solar flux in each filter (Willmer 2018), and was, then, converted to apparent albedo via normalization to a Lambert phase function at the corresponding planetary phase angle (Pallé et al. 2003; Qiu et al. 2003; Ryan & Robinson 2022). Also shown in these figures are predictions for Earth's phase-dependent apparent albedo from the Virtual Planetary Laboratory 3D spectral Earth model (Tinetti et al. 2006; Robinson et al. 2011). Because appropriate cloud and trace gassounding data from 1990 and 1992 are not available, the model predictions instead come from full-orbit simulations of Earth presented in Robinson et al. (2010).

Figure 6 shows a color–color plot for the ratio of the RED/VIOLET and IR-7560/VIOLET filter pairs. For either given pair of SSI filters, a larger ratio implies a redder average color while a smaller ratio implies a bluer average color. A colorbar is provided that maps the data points to the subspacecraft west longitude and associated key geographical features. The largest redward deviations occur when the Sahara (near the Prime Meridian) and Amazon rainforest are in view. Alternatively, the largest blueward deviation occurs when the Pacific Ocean is in view.

Finally, while data sets containing full-disk images of the distant Earth are very uncommon, they are not entirely unique to Galileo. Livengood et al. (2011) performed a very similar study on data obtained by NASA's EPOXI mission, an extension of the *Deep Impact* mission (Meech 2002). Their primary products are disk-integrated multiwavelength lightcurves (like those presented here) and time-averaged near-infrared spectra. Figure 7 compares spectrophotometry from the 1990 Galileo Earth encounter to those from the "EarthObs1," "EarthObs4," and "EarthObs5" *EPOXI* data sets, with respective phase angles of 57°.7, 75°.1, and 76°.4.

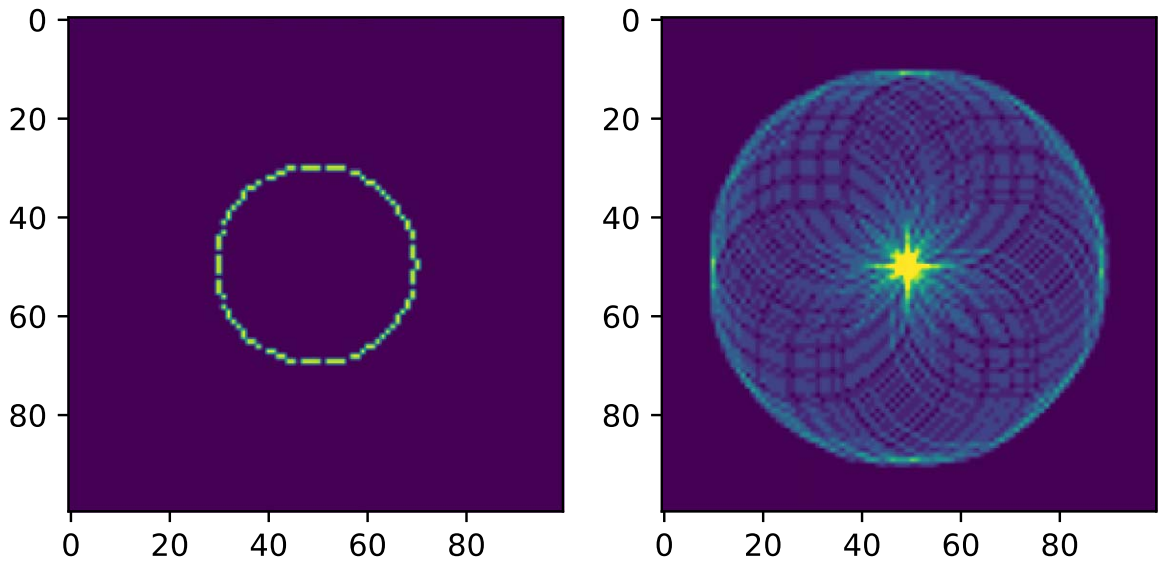

**Figure 2.** Demonstration of a two-dimensional circle Hough transform with a fixed radius. The left figure shows a synthetically generated image of a circle. The right figure shows the same image but transformed into Hough space. For each nonzero pixel on the left image, a circle is generated in the Hough transform space centered on that pixel and added to an accumulator array. The densest region within this accumulator array, shown by the bright patch at the center of the right figure, is the best-fit pixel position of the center of the circle. The dimensionality of a Hough transform can be arbitrarily high, but the known angular radius of Earth in each SSI helps to significantly constrain the approach.

## 4. Discussion

The Galileo/SSI photometry provides a unique and valuable addition to the limited existing disk-integrated observations of Earth. When viewed without prior context, the spectra (from the 1990 flyby) and lightcurves (from both the 1990 and 1992 flybys) indicate a world that, in most cases, aligns with our understanding of Earth gathered from previous analyses (especially from EPOXI). Spectra indicate a blue world (Figure 5), and lightcurves (Figures 3 and 4) demonstrate variability due to rotation (i.e., continents, oceans, and cloud structures rotating in and out of view) and time-evolving weather. Time-dependent color–color information (Figure 6) show marked deviations along a blue-red color axis, where blueward shifts have been previously attributed to Rayleigh scattering over a dark, gray surface (e.g., the Pacific Ocean Cowan et al. 2011) and redward shifts are attributed to land masses (Cowan et al. 2009).

The spectral placement and widths of the SSI filters impact qualitative inferences about the planetary environment, especially as compared to the EPOXI results. The EPOXI/HRIVIS filters are relatively wide (typically spanning 0.1 $\mu$m) and span the ultraviolet-visible range near-completely. As a consequence, the EPOXI spectrophotometric data (Figure 7) capture an increase in reflectivity across 0.7–0.9 $\mu$m (where land features and vegetation have associated increases in reflectivity) and also capture a clear drop in reflectivity in the 0.90–0.97 $\mu$m filter (attributed to water vapor absorption).

By contrast, the SSI filters are generally ∼2–4 times less wide than the HRIVIS filters and are, thus, more sensitive to narrower features in Earth's spectrum. The SSI spectrophotometry shows roughly flat/constant reflectivity in all filters except VIOLET, implying reddening due to continents and vegetation, as well as water vapor absorption, are not plainly apparent. Given the added context of knowing Earth's atmospheric composition a priori, it is likely the case that reflectivity in the narrower IR-7270 and IR-7560 filters is reduced by weak water vapor absorption and strong molecular oxygen absorption, respectively. The IR-9680 filter happens to catch a portion of the 0.95 $\mu$m water vapor band where the planetary reflectivity is very similar to that at optical wavelengths. Additionally, continent and vegetation signatures will be reduced in the disk integration process, which presents a challenge not encountered when using spatially resolved SSI data (as in Sagan et al. 1993). A full retrieval analysis could reveal if the IR-7560 and IR-9680 data provide sufficient evidence for the detection of molecular oxygen, water vapor, and/or continents/vegetation. Of course the SSI filters were not selected for the study of Earth-like environments, but their application to Earth highlights the importance of thoughtful planning for spectral and/or photometric coverages of future exoplanet direct imaging missions.

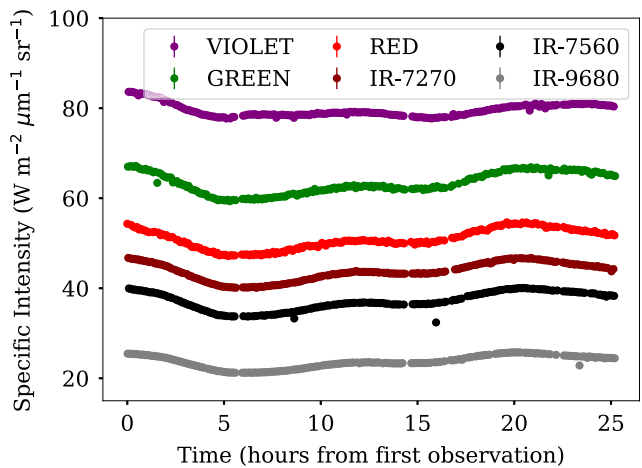

**Figure 3.** Disk-integrated specific intensity lightcurves of Earth from the 1990 Galileo encounter in the six available SSI filters.

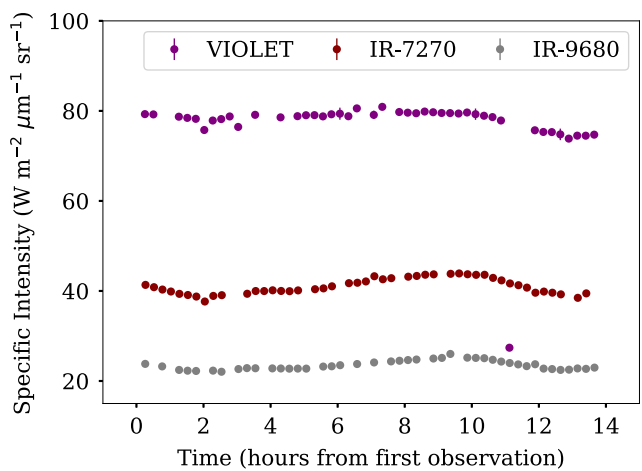

**Figure 4.** Same as Figure 3, but for the 1992 encounter.

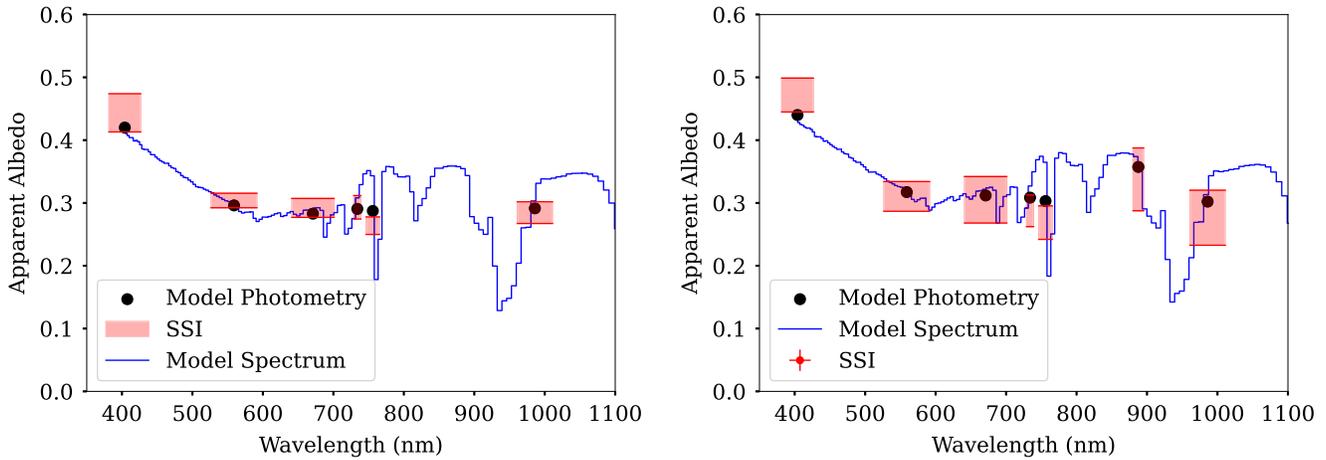

**Figure 5.** Time-averaged spectrophotometry from the 1990 (left) and 1992 (right) Galileo Earth encounters. Data from 1990 are averaged across a full rotation while data from 1992 can only be averaged over the 14 hr of available useful observations. Noisy, single-image data from 1992 are shown that were too sparse to include in lightcurve results. The shaded regions indicate the overall variation in each filter across the full encounter. We do not report photometric error bars, as the rotational change in apparent albedo is a far more dominant source of variation. Also shown are spectral and photometric predictions from the Virtual Planetary Laboratory 3D spectral Earth model, taken from phase-dependent Earth simulations presented in Robinson et al. (2010).

While the narrower bandpasses of Galileo/SSI could provide an advantage over those of EPOXI/HRIVIS for detecting sharper molecular features, there are some potential advantages of the EPOXI mission over that of Galileo. Early during the mission, the Galileo high-gain antenna failed to deploy, which severely limited the data transfer bandwidth to Earth (Johnson 1994). Information was then lost during the requisite compression of data for transmission to Earth. This, coupled with the relatively immature state of CCD technology at the time of SSI's development, likely contribute additional digitization and readout noise. In addition, many of the original calibration files for SSI are not presently available for download on NASA's PDS, making it difficult to recalibrate the Galileo data using more modern reduction pipelines. Thus, the EPOXI data provides better signal-to-noise than is possible with Galileo/SSI.

Peak-to-trough variability for the 1990 flyby lightcurves for the VIOLET, GREEN, RED, IR-7270, IR-7560, and IR-9680 filters are, respectively, 15%, 12%, 15%, 15%, 21%, and 19%. Although observed at a larger phase angle (57.°7), the EPOXI "EarthObs1," "EarthObs4," and "EarthObs5" data sets provide useful comparison points. Specifically, over a full Earth rotation data in the HRIVIS 450 nm filter (more analogous to the SSI VIOLET filter) showed 12% variability, the 550 nm filter (more analogous to the SSI GREEN filter) showed 12% variability, the 650 nm filter (more analogous to the SSI RED filter) showed 14% variability, and the 950 nm filter (which strongly overlaps the SSI IR-9680 filter) showed 19% variability. Thus, variability measures are comparable across these two experiments, which indicates that gibbous-phase rotational variability of Earth is relatively constant accross longer timescales. Notably, while the rotationally averaged reflectivity of Earth in the SSI IR-7270, IR-7560, and IR-9680 filters did not straightforwardly indicate the presence of continents and/or vegetation, the larger rotational variability in these filters (more driven by land masses rotating in/out of view than at shorter wavelengths) would provide potential evidence for continents. Studies of time-varying color in the SSI data would help to clarify this point (Cowan et al. 2009).

The Virtual Planetary Laboratory 3D spectral Earth model (Tinetti et al. 2006; Robinson et al. 2011; Schwieterman et al. 2015) provides an accurate reproduction of the rotationally averaged SSI photometry (Figure 5), although this is not particularly surprising given previous validation efforts against EPOXI photometry at these wavelengths. Nevertheless, the SSI comparisons provide further evidence that the high-fidelity Virtual Planetary Laboratory model is a suitable stand-in for Earth in mission development efforts related to exo-Earth direct imaging. Future spectral observations of Earth from spacecraft at ultraviolet/optical/near-infrared wavelengths—especially at resolving powers at or equal-to those planned for future direct imaging missions—would provide even stronger validations.

Finally, the rotationally averaged SSI photometry presents an opportunity to understand how incomplete information impacts an inferred understanding of the planetary environment, especially within the context of a future exo-Earth direct imaging mission. For example, Robinson & Salvador (2022) applied remote sensing techniques to EPOXI disk-integrated Earth data to test retrieval approaches for exoplanet direct imaging. A similar application of reflected-light retrieval tools for exoplanets (e.g., Lupu et al. 2014; Marley et al. 2014; Feng et al. 2018) to the Galileo/SSI Earth data presented here is warranted and is a clear next step.

## 5. Conclusions

The Galileo encounters with Earth in 1990 and 1992 provided much richer data sets than had been previously reported (Sagan et al. 1993). Work presented here developed a pipeline to vet and calibrate exoplanet analog images from Galileo/SSI, resulting in over 1500 useable photometric measurements. For the 1990 encounter, data span more than a full Earth rotation in six (of eight available) SSI filters and provide rotational lightcurves in each of these filters. Data from the 1992 encounter are lower-quality, with 14 hr lightcurves reduced in three filters and noisy, single-pointing photometry reduced in four other filters. The time-averaged photometry reported here from these two data sets are a significant contribution to the very limited number of published data sets measuring disk-integrated lightcurves for Earth at phase angles appropriate for exoplanet direct imaging. The variability and color measurements reported here are consistent with related findings from the EPOXI mission. Reflectance measured in the narrow IR-7560 SSI filter could indicate absorption due to the molecular oxygen $A$-band feature at 760 nm. Finally, comparisons to predictions from the Virtual Planetary Laboratory 3D spectral Earth model serve to further validate this tool as a useful stand-in for Earth in the continued absence of high-quality, disk-integrated spectroscopic observations of our planet across a broad range of ultraviolet and optical wavelengths.

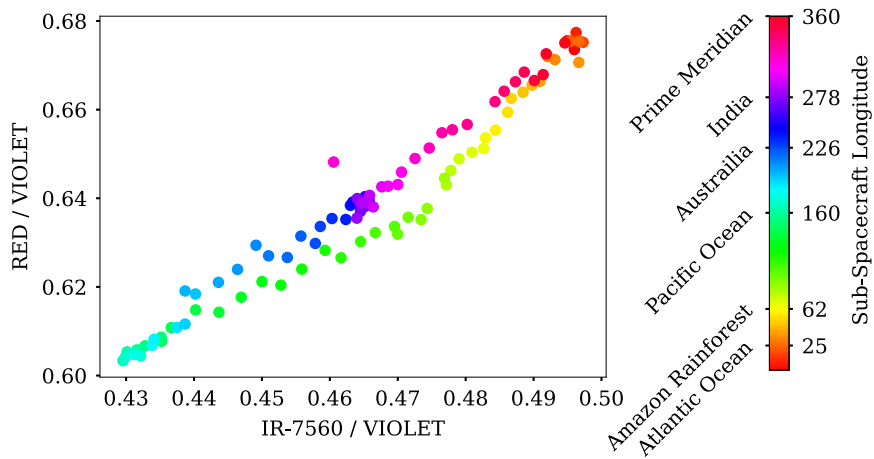

**Figure 6.** Color–color plot from the 1990 Galileo encounter with Earth showing brightness ratios in the RED/VIOLET and IR-6560/VIOLET SSI filters. The colorbar on the right indicates sxub-spacecraft west longitude, with important features marked along that colorbar. Larger values along either axis indicate redder colors. The overall color of Earth changes in a cyclical fashion as it rotates underneath the spacecraft. The strongest redward deviation occurs when the Sahara and Amazon rainforest were in view while the strongest blueward deviation occurs when the Pacific Ocean is in view.

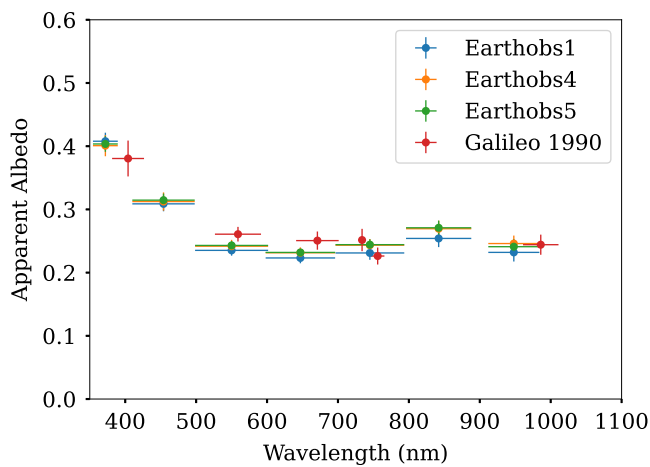

**Figure 7.** Comparison between the time-averaged reflectance spectrum from the Galileo 1990 encounter with Earth and that obtained from the EarthObs1, EarthObs4, and EarthObs5 EPOXI encounters (Livengood et al. 2011). The horizontal bars indicate the respective filter widths, while the vertical bars represent the variation in each filter across each encounter. The overall shape of the spectra are all in excellent agreement. While EPOXI does have broader bandpasses and better signal-to-noise, Galileo/SSI can capture narrower features that are smoothed over in the wider EPOXI bandpasses.

## Acknowledgments

R.H.S. and D.E.T. acknowledge support from NASA award No. 80NSSC20K0670. T.D.R. gratefully acknowledges support from NASA's Exoplanets Research Program (No.80NSSC 18K0349), Habitable Worlds Program (No.80NSSC20K0226), Exobiology Program (No. 80NSSC19K0473), the Nexus for Exoplanet System Science Virtual Planetary Laboratory (No. 80NSSC18K0829), and the Cottrell Scholar Program administered by the Research Corporation for Science Advancement. This research has made use of the USGS Integrated Software for Imagers and Spectrometers (ISIS).

*software:* USGS Integrated Software for Imagers and Spectrometers (ISIS; Laura et al. 2023); Astropy Collaboration (2022); Scikit-Image (van der Walt et al. 2014).

## ORCID iDs

Ryder H. Strauss  https://orcid.org/0000-0001-6350-807X
Tyler D. Robinson  https://orcid.org/0000-0002-3196-414X
David E. Trilling  https://orcid.org/0000-0003-4580-3790